\documentclass[runningheads,a4paper]{llncs}

\usepackage{amssymb}
\usepackage{graphicx}
\usepackage{epstopdf}
\usepackage{graphics}

\usepackage{geometry}
\usepackage{epsfig}
\usepackage{subfigure}
\usepackage{amsmath}
\usepackage{graphics}
\usepackage{amssymb}
\usepackage{alltt}
\usepackage{float}
\usepackage{array}
\usepackage{xspace}
\usepackage{footmisc}
\usepackage{comment}
\usepackage{algorithm}
\usepackage{algorithmic}

\usepackage{url}
\usepackage[justification=centering]{caption}

\usepackage{url}
\urldef{\mailsa}\path|{alfred.hofmann, ursula.barth, ingrid.haas, frank.holzwarth,|
\urldef{\mailsb}\path|anna.kramer, leonie.kunz, christine.reiss, nicole.sator,|
\urldef{\mailsc}\path|erika.siebert-cole, peter.strasser, lncs}@springer.com|
\urldef{\mailsd}\path|{jlong, leizhu, cyzhang, linshq, zyang22}@csu.edu.cn, xpyuan@hut.edu.cn|
\newcommand{\keywords}[1]{\par\addvspace\baselineskip
\noindent\keywordname\enspace\ignorespaces#1}

\begin{document}

\mainmatter  

\title{HOC-Tree: A Novel Index for efficient Spatio-temporal Range Search}

\titlerunning{HOC-Tree: A Novel Index for efficient Spatio-temporal Range Search}

%
%
\author{}
\authorrunning{HOC-Tree: A Novel Index for efficient Spatio-temporal Range Search}

\author{ Jun Long$^\dagger$, Lei Zhu$^\dagger$, Chengyuan Zhang$^\dagger$, Shuangqiao Lin$^\dagger$, Zhan Yang$^\dagger$, Xinpan Yuan$^\ddagger$}

\institute{$^\dagger$ School of Information Science, Central South University, PR China\\
$^{\ddagger}$ School of Computer, Hunan University of Technology, China\\
\mailsd\\
}

%
%

\toctitle{Lecture Notes in Computer Science}
\tocauthor{Authors' Instructions}
\maketitle

\begin{abstract}
With the rapid development of mobile computing and Web services, a huge amount of data with spatial and temporal information have been collected everyday by smart mobile terminals, in which an object is described by its spatial information and temporal information. Motivated by the significance of spatio-temporal range search and the lack of efficient search algorithm, in this paper, we study the problem of spatio-temporal range search (STRS), a novel index structure is proposed, called HOC-Tree, which is based on Hilbert curve and OC-Tree, and takes both spatial and temporal information into consideration. Based on HOC-Tree, we develop an efficient algorithm to solve the problem of spatio-temporal range search. Comprehensive experiments on real and synthetic data demonstrate that our method is more efficient than the state-of-the-art technique.
\end{abstract}

\section{Introduction}
\label{intro}

With the rapid development of mobile computing and Web services, a huge amount of data~\cite{DBLP:journals/tip/WangLWZ17,DBLP:journals/tip/WangLWZZH15,DBLP:journals/tnn/WangZWLZ17,DBLP:conf/mm/WangLWZZ14} with spatial and temporal information have been collected everyday by smart mobile terminals, such as smart phones, tablets, wearable devices etc, or devices of Iot which are equiped with GPS or wireless modules. In addition, Location Based Services (LBS) and social network services provide users with location-dependent information and services in dailylife. Everyday a vast number of pictures~\cite{DBLP:journals/pr/WuWLG18,DBLP:journals/cviu/WuWGHL18,DBLP:conf/sigir/WangLWZZ15} and texts with geotags~\cite{DBLP:journals/tkde/ZhangZZL16,DBLP:conf/icde/ZhangZZL13,DBLP:conf/edbt/ZhangZZLCW14}and timestamps are posted to Fackbook or Instagram. Foursquare supports more than 45 million users who have checked-in more than 5 billion times at over 1.6 million businesses. Users can search any interested information by specified time interval and geolocation.

\begin{figure}[thb]
\newskip\subfigtoppskip \subfigtopskip = -0.1cm
\centering
\includegraphics[width=.60\linewidth]{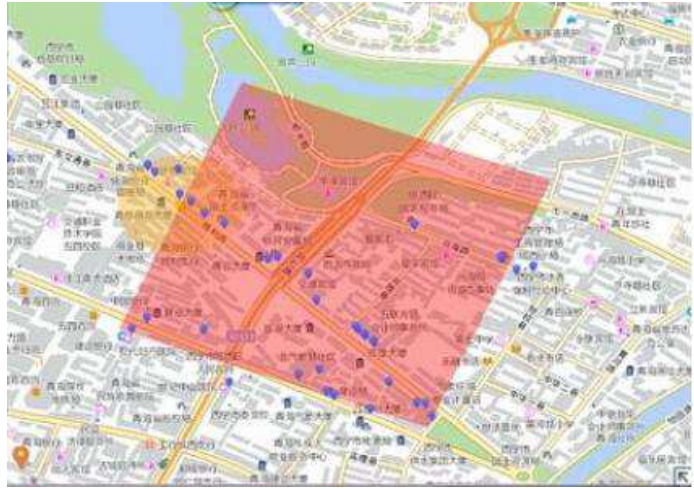}
\vspace{-1mm}
\caption{\small an example of spatio-temporal search in ocation-based social network services }
\label{fig:Fig1}
\end{figure}

In this paper, we study an important search problem in spatio-temporal data query area, named spatio-temporal range search (STRS for short). Spatio-temporal range search aims to retrieval all spatio-temporal objects whose location is within a specific geographical region during a time region. In many application scenarios, it plays an important role for data management and geo-social networks. For example, in location-based social networks platforms, such as Facebook, Twitter, Weibo, etc.. Users prefer to make friends with the people who usually do daily activities in the same geographic region and same time range, because same daily activities like shopping, doing outdoor exercises, going to cinema, etc. are important factors to establish relationships. Thus according to the posts with spatio-temporal data, they can find the users who have the same hobbies within a given area and given time interval, shown in Fig.~\ref{fig:Fig1}. The red square is the geographical range of search for daily activities. Likewise, location based services like Facebook’s Nearby and Foursquare’s Radar return the friends that recently checked-in at close proximity to a user’s current location~\cite{DBLP:journals/pvldb/ArmenatzoglouPP13,DBLP:journals/tip/WangLWZ17,WangINF,LinWangIVC,YangIJCAI16,DBLP:journals/pr/WuWGL18,NNLS2018}. In the big data age, as swift growth of the amount of spatio-temporal data, spatio-temporal range search has become a hot issue in data searching and management area.

\subsubsection{Motivation.}
 The challenges for the problem of spatio-temporal range search are two-fold. (i) Due to the massive amount of spatio-temporal objects in lots of important applications, large-scale heterogeneous social networks with spatial and temporal information have been constructed. How to efficient management and access geo-social data is a core problem. (ii) For the various application requirements in social network services, high efficient search algorithms need to be developed by combining spatial and temporal features of social data.

Motivated by the significance of spatio-temporal range search and the lack of efficient search algorithm, we propose a novel spatio-temporal index structure, named HOC-Tree based on Hilbert curve and OC-Tree. Besides, we develop an efficient range search algorithm for STRS problem. OC-Tree is an important index structure in spatial database area. It is most often used to partition a three-dimensional space by recursively subdividing it into eight octants. HOC-Tree is a nature extension of OC-Tree, but it not only inherits the valuable properties in 3-dimensional partition, such as the data that close in space and close in time are partition into same cell, but also provides an efficient 3D Morton code generation mechanism, which can easily and effectively combine the spatial and temporal information together to support spatio-temporal search.

\subsubsection{Contributions.}
 To summarize, our key contributions in this paper are summarized as follows:

(1) We propose a novel spatio-temporal index based on Hilbert curve and OC-Tree named HOC-Tree to solve the problem of spatio-temporal range search. To the best of our knowledge, this study is the first time to design a novel spatio-temporal indexing mechanism for efficient spatio-temporal range search.

(2) We develop an efficient spatio-temporal range search algorithm based on HOC-Tree.

(3) We conduct comprehensive experiments on real and synthetic datasets. The results show that our method can solve spatio-temporal range search effectively and efficiently, and it outperforms the state-of-the-art approaches.

\subsubsection{Roadmap.}
 The rest of the paper is organized as follows: We present the related work in Section 2. Section 3 formally defines the problem and describes the index structure. We elaborate the search algorithm in Section 4 and extensive experiments are presented in Section 5. Finally, we offer conclusions in Section 6.

\section{Related work}
\label{related}
In this section, we review geo-social networks queries and collective spatial queries，which are two kinds of techniques related to our works.

\subsubsection{Geo-social networks queries.}
A typical geo-social network ~\cite{DBLP:journals/corr/abs-1708-02288,TC2018,DBLP:conf/mm/WangLWZZ14} combines social networks techniques ~\cite{DBLP:conf/pakdd/WangLZW14,DBLP:journals/pr/WuWGL18,WangMM15} and spatial data queries techniques. Many research findings from academia and techniques applying to industry have been proposed. In industrial circle, the most famous social networks platform, Facebook, provided a location-based social network service named \emph{Nearby}~\cite{DBLP:journals/pvldb/ArmenatzoglouPP13} which aims to find the friends who are in the neighborhood of a user currently. Geoloqi is another analogous platform for building location aware applications. It provides the service which notifies users when their friends get into a certain geographical region. Uber is a advanced mobile Internet platform for texi service based on geo-location information of texi drivers and riders. The riders can search the near drivers around them and send messages for request. These applications just only focus on the spatial attributes of data on social networks or cloud platforms for range search. In academic circles, the problem of spatio-temporal search is concerned by lots of researchers. In~\cite{DBLP:journals/pvldb/ArmenatzoglouPP13}, armenatzoglou el at. proposed a general framework that offers flexible data management and algorithmic design. The nearest star group query contained in the framework returns the \emph{k} nearest subgraphs of \emph{m} users. In~\cite{DBLP:conf/dasfaa/LiuSCHJC12}, Liu et al. proposed propose the \emph{k}-Geo-Social Circle of Friend Query which aims to finds the group g of \emph{k} + 1 users, which (i) is connected, (ii) contains \emph{u}, and (iii) minimizes the maximum distance between any two of its members. In~\cite{DBLP:conf/icwsm/ScellatoNLM11}, Scellato et al. proposed three more geo-social networks metrics: (i) average distance (ii) distance strength, and (iii) average triangle length. In~\cite{DBLP:conf/kdd/YangSLC12}, Yang et al. developed a hybrid index named Social R-Tree to solve the problem of socio-spatial group query. The studies mentioned above did not combine the spatial and temporal attributes of objects in database for searching.

 \subsubsection{Geo-social networks queries.}
 Collective spatial query is another important problem. In~\cite{DBLP:conf/icde/ZhangCMTK09}, Zhang et al. presented a novel spatial keyword query problem called the \emph{m}-closest keywords (\emph{m}CK) query, which aims to aims to find the spatially closest tuples which match \emph{m} user-specified keywords. They proposed a new index named bR*-tree extended from the R*-tree to address this problem. The R*-tree designed by Beckmann et al., which incorporates a combined optimization of area, margin and overlap of each enclosing rectangle in the directory~\cite{DBLP:journals/dr/Sellis00}. In~\cite{DBLP:conf/sigmod/GuoCC15}, Guo et al. proved that answering \emph{m}CK query is NP-hard and designed two approximation algorithms called \emph{SKEC}a and \emph{SKEC}a+. In~\cite{DBLP:journals/tkde/DengLLZ15}, Deng et al. proposed a generic version of closest keywords search named best keyword cover which considers inter-objects distance as well as the keyword rating of objects. In~\cite{DBLP:conf/sigmod/LongWWF13}, Long et al. studied two types of the CoSKQ problems, MaxSum-CoSKQ and Dia-CoSKQ. These studies aim to solve the problem of spatial keyword queries to find a set of objects. They did not develop efficient index structure and search algorithms for range search in a specific geographical area and time interval.
\section{Model and structure}
\label{model}

This section first presents a Definition of problem, then describes the proposed data structure, named HOC-Tree, which based on Hilbert curve and OC-Tree. Table ~\ref{tab:notation} below summarizes the symbols used frequently throughout the paper.

\begin{table}
	\centering
    \small
	\begin{tabular}{|p{0.20\columnwidth}| p{0.69\columnwidth} |}
		\hline
		\textbf{Notation} & \textbf{Definition} \\ \hline\hline
		~$D$                & a given data set of spatio-temporal data                                \\ \hline
        ~$o$                & a spatio-temporal object                                \\ \hline
        ~$x_i$              & the longitude of a spatio-temporal data  \\ \hline
		~$y_i$              & the latitude of a spatio-temporal data                              \\ \hline
		~$t_i$              & the timestamp of a spatio-temporal data                                \\ \hline	
    	~$h_n$              & the value of Morton order in Hilbert curve \\\hline
        ~$L$                & the deepest level of HOC-Tree \\\hline
        ~$\psi$             & the division threshold value for a node in HOC-Tree \\\hline
        ~$v$                & the Morton order of a leaf node in HOC-Tree                               \\ \hline	
        ~$q$                & a spatio-temporal range search ([$x_{min}$, $x_{max}$], [$y_{min}$, $y_{max}$], [$t_{start}$, $t_{end}$])                              \\ \hline
        ~$d$                & the Euclidean distance between two points in spatial space                                \\\hline
	\end{tabular}
    \caption{The summary of notations} \label{tab:notation}	
\end{table}

\subsection{Problem Definition}

\begin{definition}[\textbf{Spatio-temporal Object Set}] \label{def:range stos}
A spatio-temporal object set can be defined as $D=\{o_1, o_2, …, o_n\}$. Each spatio-temporal object $o$ is associated with a spatial location $o$.$($$x_i$, $y_i$$)$ and the timestamp $o.t_i$.
\end{definition}

\begin{definition}[\textbf{Spatio-temporal Range Search (STRS)}] \label{def:range strs}
Given a spatio-temporal objects data set $D$, a range query is defined as $q$$($$[$$x_{min}$, $x_{max}$$]$, $[$$y_{min}$, $y_{max}$$]$, $[$$t_{start}$, $t_{end}$$]$$)$ where $($$[$$x_{min}$, $x_{max}$$]$, $[$$y_{min}$, $y_{max}$$]$$)$ is the query spatial region and $($$[$$t_{start}$, $t_{end}$$]$$)$ is the query temporal interval, this work aims to select all the records which satisfy the query $q$ from $D$.
\end{definition}

\subsection{Index structure}
In this section, we introduce a novel spatio-temoral index, named HOC-Tree, which is based on OC-Tree and Hilbert curve. This data structure is the key technique of this work.

\begin{figure}[thb]
\newskip\subfigtoppskip \subfigtopskip = -0.1cm
\centering
\includegraphics[width=.60\linewidth]{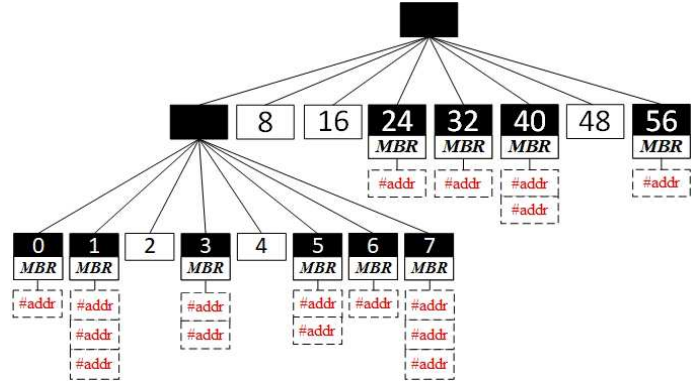}
\vspace{-1mm}
\caption{\small An example of HOC-Tree with \emph{MBRsign} tags }
\label{fig:Fig6}
\end{figure}

As it will be shown in Subsection 4.1, the more subspaces overlapping with range query \emph{q}, the more time will be consumed when searching HOC-Tree. To solve this problem, a \emph{MBRsign} tag data structure is used to reduce non-promising nodes access, which can avoid unnecessary I/O costs. For each subspace, the spatial locations of all the points in it can be associated with a minimum bounding rectangle (MBR), so a \emph{MBRsign} tag is maintained for each non-empty leaf node to keep the MBR information. For a given range query q, the covering non-empty leaf nodes which don’t satisfy the spatial constraint will not be accessed in searching process with the help of tags. HOC-Tree keeps two end points information of the MBR, which only require 16 bytes for each non-empty leaf node. The more detail of using the tags will be described particularly in Subsection 4.1, where elucidates the search algorithms. Fig.~\ref{fig:Fig6} illustrates the structure of HOC-Tree with \emph{MBRsign} tags.

The black blocks represent non-empty nodes which contains a list of spatio-temporal data locations while the white blocks represent empty nodes. Each leaf node is labeled by its Morton order value according to our approach as mentioned above, and tags are kept for them to maintain the MBR.

\section{Spatio-temporal query algorithms}
\label{algorithms}

This section gives exhaustive description of spatio-temporal range search based on HOC-Tree.

\subsection{Range search algorithm}
Range query $q$ is an essential function in spatio-temporal data processing. In our algorithms as shown in \textbf{Algorithm 1}, this work is done in several stages. The input query $q$ = $($$[$$x_{min}$, $x_{max}$$]$, $[$$y_{min}$, $y_{max}$$]$, $[$$t_{start}$, $t_{end}$$]$$)$ is in three-dimensional space, where $[$$x_{min}$, $x_{max}$$]$ gives the range of longitude, $[$$y_{min}$, $y_{max}$$]$ gives the range of latitude and $[$$t_{start}$, $t_{end}$$]$ gives the time interval. The output S is a set of entries inside spatio-temporal query $q$. This algorithm only accesses the optimized nodes when searching HOC-Tree. A prune process is executed to check the entries whether they satisfy the query range or not and remove false positives to refine results.

\begin{algorithm}
\begin{algorithmic}[1]
\footnotesize
\caption{\bf Spatio-Temporal Range Search}
\label{alg:strs}

\INPUT  $q~:x_{min}, x_{max}, y_{min}, y_{max}, t_{start}, t_{end}$.
\OUTPUT $S~: $set of entries inside query range $q$.

\STATE $S\leftarrow\emptyset$;$CoveringSpacialSpaces\leftarrow\emptyset$;$RegionSet\leftarrow\emptyset$;
\STATE $CoveringSpacialSpaces\leftarrow getHilbertValues$($x_{min}$, $x_{max}$, $y_{min}$, $y_{max}$);
\STATE $RegionSet←getRegions(CoveringSpacialSpaces)$;
\STATE $CoveringNodes\leftarrow\emptyset$; $Q.L^f\leftarrow\emptyset$; $Q.L^p\leftarrow\emptyset$;

\FOR{each region $\in$ $RegionSet$}
    \STATE $CoveringNodes\leftarrow getOverlappingCubes(x_{min}, x_{max}, y_{min}, y_{max}, t_{start}, t_{end})$;
    \STATE $N^f, N^p\leftarrow Identify(CoveringNodes)$;
        \FOR{each Node $v \in N^p$}
            \STATE $\textbf{MBRCheck}(MBRsign_v, x_{min}, x_{max}, y_{min}, y_{max})$;
        \ENDFOR
        \FOR{each Node $v \in N^p$ \emph{that survive from} $\textbf{MBRCheck}$ \emph{process}}
            \STATE $Q.L^p\leftarrow getEntriesList(v)$;
        \ENDFOR
        \FOR{each Node $v \in N^f$}
            \STATE $Q.L^f\leftarrow getEntriesList(v)$;
        \ENDFOR
        \STATE $S\leftarrow Q.L^f + Prune(Q.L^p);$
\ENDFOR
\RETURN {$S$}.
\end{algorithmic}
\end{algorithm}

\emph{\textbf{Mapping Hilbert curve Values:}} For a given range query $q$, the Hilbert curve values of covering spatial spaces can be calculated immediately according to the region $($$[$$x_{min}$, $x_{max}$$]$, $[$$y_{min}$, $y_{max}$$]$$)$ of $q$. The function \emph{getHilbertValues}() maps the rectangle region into a set of one-dimensional values in line 2.

\begin{figure}[thb]
\newskip\subfigtoppskip \subfigtopskip = -0.1cm
\centering
\includegraphics[width=.60\linewidth]{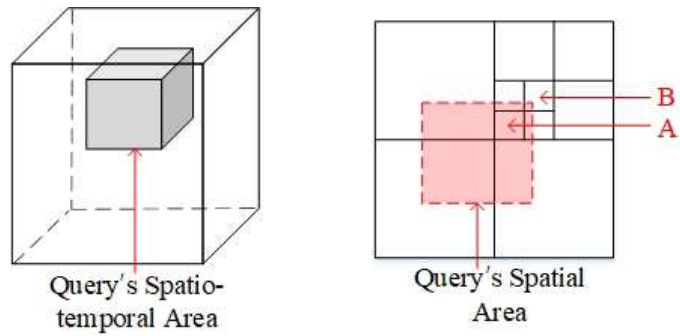}
\vspace{-1mm}
\caption{\small Spatio-temporal range search }
\label{fig:Fig7}
\end{figure}

\textbf{\emph{Finding Spatio-temporal Covering Cubes:}} Before searching HOC-Tree in corresponding regions locally, the function \emph{getOverlappingCubes}() line in 6 computes the covering nodes which overlap with three-dimensional query range. The covering cubes can be partial or full. The left part of Fig.~\ref{fig:Fig7} shows a spatio-temporal range query $q$ (the shaded cube) which would overlap multiple subspaces. For simplicity, the partition of space does not present here. The cubes overlapping with query range in spatial dimension is illustrated in the right part of Fig.~\ref{fig:Fig7}, where the deepest level $L$ of the HOC-Tree is 4. We can see that cube A has a full spatial overlap while the rest cubes have partial spatial overlap.

For each covering node, it needs to be searched HOC-Tree to get the list of addresses refer to the locations of data point. All the points in fully spatio-temporal overlapping cubes will satisfy the spatio-temporal range search which do not need to do an additional refinement step. \textbf{Algorithm 2 Identify} distinguishes these two kinds of covering nodes by $N^f$ and $N^p$, where $N^f$ denotes the set of fully covering nodes and $N^p$ denotes the set of partially covering nodes. The identification of full overlaps helps to reduce the computation time, which can avoid unnecessary CPU checking overhead in the refinement step.

\begin{algorithm}
\begin{algorithmic}[1]
\footnotesize
\caption{\bf Identify(CoveringNodes)}
\label{alg:strs}
\INPUT  $CoveringNodes$.
\OUTPUT $N^p,N^f$

\STATE $N^f\leftarrow\emptyset$;$N^p\leftarrow\emptyset$;
\FOR{each region $\in$ $CoveringNodes$}
    \IF{$v$ is a full overlapping leaf node}
        \STATE add $v$ into $N^f$;
    \ENDIF
    \IF{$v$ is a partial overlapping leaf node}
        \STATE add $v$ into $N^p$;
    \ENDIF
\ENDFOR
\RETURN {$N^p,N^f$}.
\end{algorithmic}
\end{algorithm}

\begin{figure}[thb]
\newskip\subfigtoppskip \subfigtopskip = -0.1cm
\centering
\includegraphics[width=.60\linewidth]{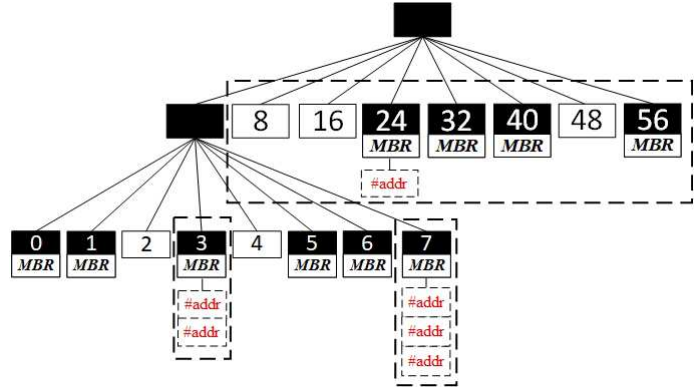}
\vspace{-1mm}
\caption{\small Query overlapping nodes and non-empty covering nodes }
\label{fig:Fig8}
\end{figure}

\emph{\textbf{Confirming Non-empty Covering Nodes:}} The benefit of coupling the spatial and temporal information in our index will be more clear in this stage. As shown in Fig.~\ref{fig:Fig7}(right part), the overlap of cube B with query’s spatial dimensional area is very small w.r.t cube A, which has a full overlap. If searching the index without any spatial discrimination, then a very small overlap (i.e., the cube B) will need the same I/O costs with that of a full overlap (i.e., the cube A). As a result, many false positive results will be collected, which have to be later pruned through the spatial criteria. Especially when the data is skew, there might be a lot of empty partial covering nodes. This case can happen because the points in that cube do not satisfy with the spatial criteria of $q$. In line 8 to 10, the information kept in \emph{MBRsign} tag is used to check whether the MBR overlap with the spatial criteria or not. The checking is just needed in partial covering nodes because the points in full overlaps will all satisfy the spatio-temporal criteria. The confirmation of non-empty spatial covering nodes can efficiently reduce the number of false positive results in region search. As shown in Fig ~\ref{fig:Fig8}, the Morton values of overlapping nodes (i.e. the nodes in the rectangle marked with dotted lines) is obtained by the function \emph{getOverlappingCubes}(), which overlap with the range query given in Fig.~\ref{fig:Fig7}. For simplicity, the further division of the Node $v_3$ as shown is omitted here. Then with the help of \emph{MBRsign} tag, it can further confirm the non-empty covering nodes (i.e. the Node $v_3$, the Node $v_7$ and the Node $v_{24}$), which need to be searched in HOC-Tree.

\emph{\textbf{Searching HOC-Tree:}} Spatio-temporal adjacent nodes will be stored nearby each other by this encoding in HOC-Tree. Identifying full and partial covering nodes helps to reduce the computation time, while confirming non-empty spatial covering nodes can reduce the number of I/O during searching HOC-Tree. Furthermore, the property of Hilbert curve can ensure that the generated Morton value of query range will contain all the valid points, which discussed in Subsection 3.3. According to the stages described above, the algorithm can get a set of full covering nodes $N^f$ and a set of non-empty partial covering nodes $N^p$ which need to be searched in HOC-Tree. Line 11 to 16 give the search result by $Q.L^f$ and $Q.L^p$ , where the notation $Q.L^f$ and $Q.L^p$ denote the sets of entries in full and partial covering nodes respectively.

\emph{\textbf{ Refining Results:}} The entries in $Q.L^p$ which have partial overlap need further refinement. There might be some points in partial overlapping nodes that not satisfy with the spatio-temporal query range. \textbf{Algorithm 3 Prune} checks each entry in $Q.L^p$ whether it is inside query range or not and removes unrelated results immediately.

\begin{algorithm}
\begin{algorithmic}[1]
\footnotesize
\caption{\bf Prune($Q.L^p$)}
\label{alg:prune}
\INPUT  $Q.L^p$
\OUTPUT $results$

\STATE $Results\leftarrow\emptyset$;
\WHILE{$Q.L^p \ne \emptyset$}
    \FOR{each entry $ e \in Q.L^p$}
        \IF{query range contains $e$}
            \STATE add $e$ into results;
        \ENDIF
    \ENDFOR
\ENDWHILE
\RETURN {$results$}.
\end{algorithmic}
\end{algorithm}

\section{Experiments}
\label{algorithms}

\subsection{Experiment settings and datasets}
With the implementation of HOC-Tree, a comprehensive experimental evaluation is conducted to verify the performance of the scheme in a real cloud environment. The locations of all datasets were scaled to the two-dimensional space $[0, 10000]^2$, and the timestamp of all datasets were scaled to $[0, 5000]$. In addition, the spatial region grew from $200*200$ to $1000*1000$, the time interval varied from 200 to 1000, and $k$ changed from 10 to 500. By default, spatial region, time interval and $k$ were set to \textbf{600}, \textbf{600}, \textbf{100} respectively. We conducted experiments on a 3GHz Intel Core i5 2320 CPU and 8GB RAM running 64-bit Ubuntu 16.04 LTS.

Three different datasets were used in the experiments, one of which was a synthetic uniform dataset (UN) generated by program, and others were real-world datasets, described as following: the first one was collected in Geolife project [22] (GL) by 182 users from April 2007 to August 2012, the second one was T-Drive [23,24] (TD) generated by 33 thousand taxis on Beijing road network over a period of 3 months.

For accurate analyzing and evaluating, STEHIX was chosen as comparative object, which has a similar index scheme with ours. In each experimental case, the process was repeated for 5 times and the average value was reported. For the HOC-Tree in all the tests, the deepest level $L$ was set to 16 and the division threshold value $\psi$ was set to 200.

\subsection{Performance evaluation}
\subsubsection{Evaluation on different datasets.}
A series of evaluation was performed on index construction time, index size and data query performance separately against three datasets GL, TD and UN, where other parameters had default settings.

Figs depicts the rate of space occupying by the index sizes. STEHIX requires more space due to the two kinds of indices (called \emph{s-index} and \emph{t-index}) kept for all the entries. The storage cost of the STEHIX increases faster in larger datasets. In contrast, an index record is maintained for each entry only once so that our index saves more space in memory. Particularly, HOC-Tree with \emph{MBRsign} tag occupies a very small index size compared with HOC-Tree without \emph{MBRsign} tag. Figs show the difference of construction time between HOC-Tree and STEHIX. Due to the simple split and code algorithms of an HOC-Tree, our method has a shorter constructing time as compared to STEHIX, which need to traverse two indices during the construction.

\begin{figure}[thb]
\newskip\subfigtoppskip \subfigtopskip = -0.1cm
\centering
\includegraphics[width=.60\linewidth]{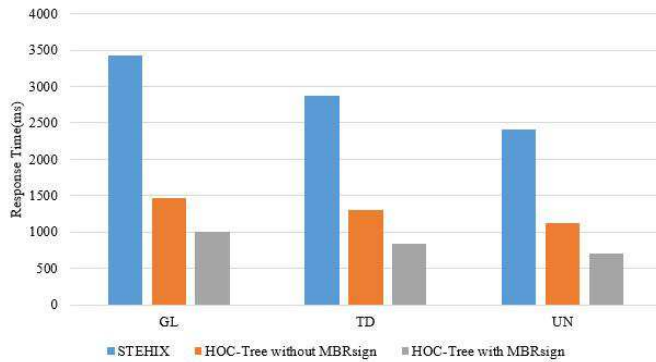}
\vspace{-1mm}
\caption{\small Data query performance on various datasets }
\label{fig:Fig12}
\end{figure}

The experiment results of range query on different datasets are shown in Fig.~\ref{fig:Fig12}, where the spatial region and time interval were both set 600. The query performance was measured by computing the duration time between when the regions started searching and when client received all accurate results.

The HOC-Tree demonstrates superior performance in comparison with STEHIX. Our analysis is as follows, STEHIX calculates the number of addresses in \emph{s-index} and \emph{t-index} separately to choose the high-selectivity list for further retrieval. In sense that, each query will decompose into two processes to collect results in temporal dimension and spatial dimension, which will provoke more I/O costs. On the other hand, STEHIX uses a period time $T$ to divide all entries in temporal dimension because of the periodicity in timestamps, which means if let $T$ = 24 (a period of 24 hours is a cycle) and divide $T$ into several segments such as 8 segments, then all the entries will map into the 8 segments by their temporal information. For a given temporal range query, all results returned by STEHIX are confused by modulo value, which have same time intervals but different in dates. Therefore, it will take more time to remove false positive results, which delays the response time in queries.

HOC-Tree with \emph{MBRsign} tag improves a little efficiency comparing with HOC-Tree without \emph{MBRsign} tag for uniform data (UN), because the benefit of MBR information maintained in \emph{MBRsign} is apparently in skew data such as GL and TD.

Apparently, larger spatial region means larger spatial search area, which results in longer response time. Therefore, both of two indexes perform better when the spatial region is small. On the other hand, the performance of range query in uniform dataset is better than real-world dataset, which mainly because that real-world dataset is a skew data. As is evident from the experiments, HOC-Tree shows an improvement over STEHIX especially when the spatial region is large. Larger spatial query range leads to much more unrelated entries identified as candidates in STEHIX. It spends much more running time and CPU cost because of the high computation for refinement step. However, our index performs better due to the non-fully-decoupling spatial and temporal properties so that all the points are placed by their spatio-temporal proximity in HOC-Tree which can help to reduce I/O load when searching trees. For a given three-dimensional query, HOC-Tree can immediately locate the covering nodes and explore the corresponding HOC-Tree which is owing to the efficient nodes’ pruning and the use of Morton value. As pointed out earlier, STRS identifies as more full covering nodes as possible during executing query operation, which helps to reduce the CPU cost for checking fully satisfied entries.

\begin{figure*}
\newskip\subfigtoppskip \subfigtopskip = -0.1cm
\begin{minipage}[b]{0.99\linewidth}
\begin{center}
     \subfigure[\small{Temporal interval on TD}]{
     \includegraphics[width=0.47\linewidth]{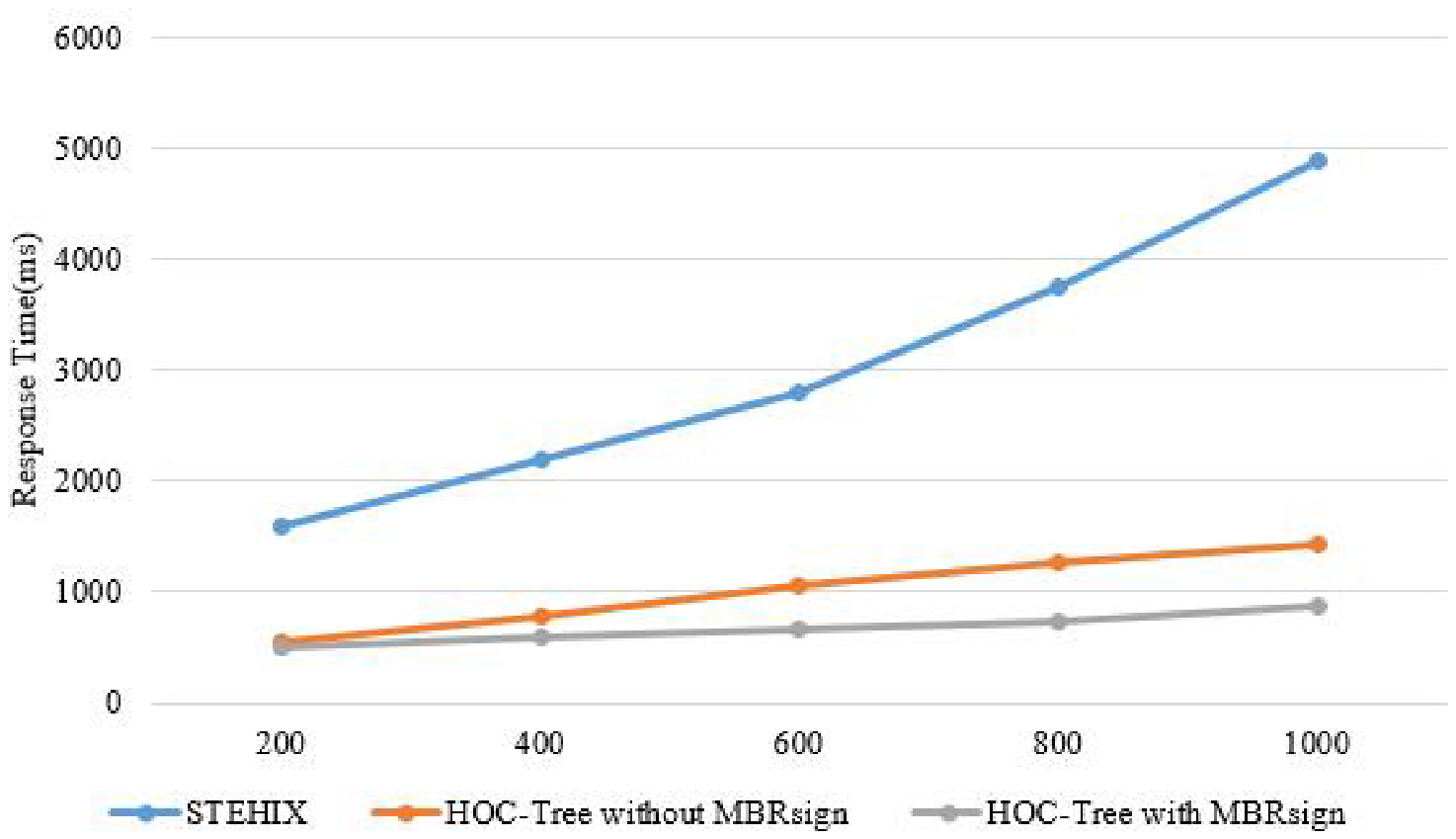}
     }
     \subfigure[\small{Temporal interval on UN}]{
     \includegraphics[width=0.47\linewidth]{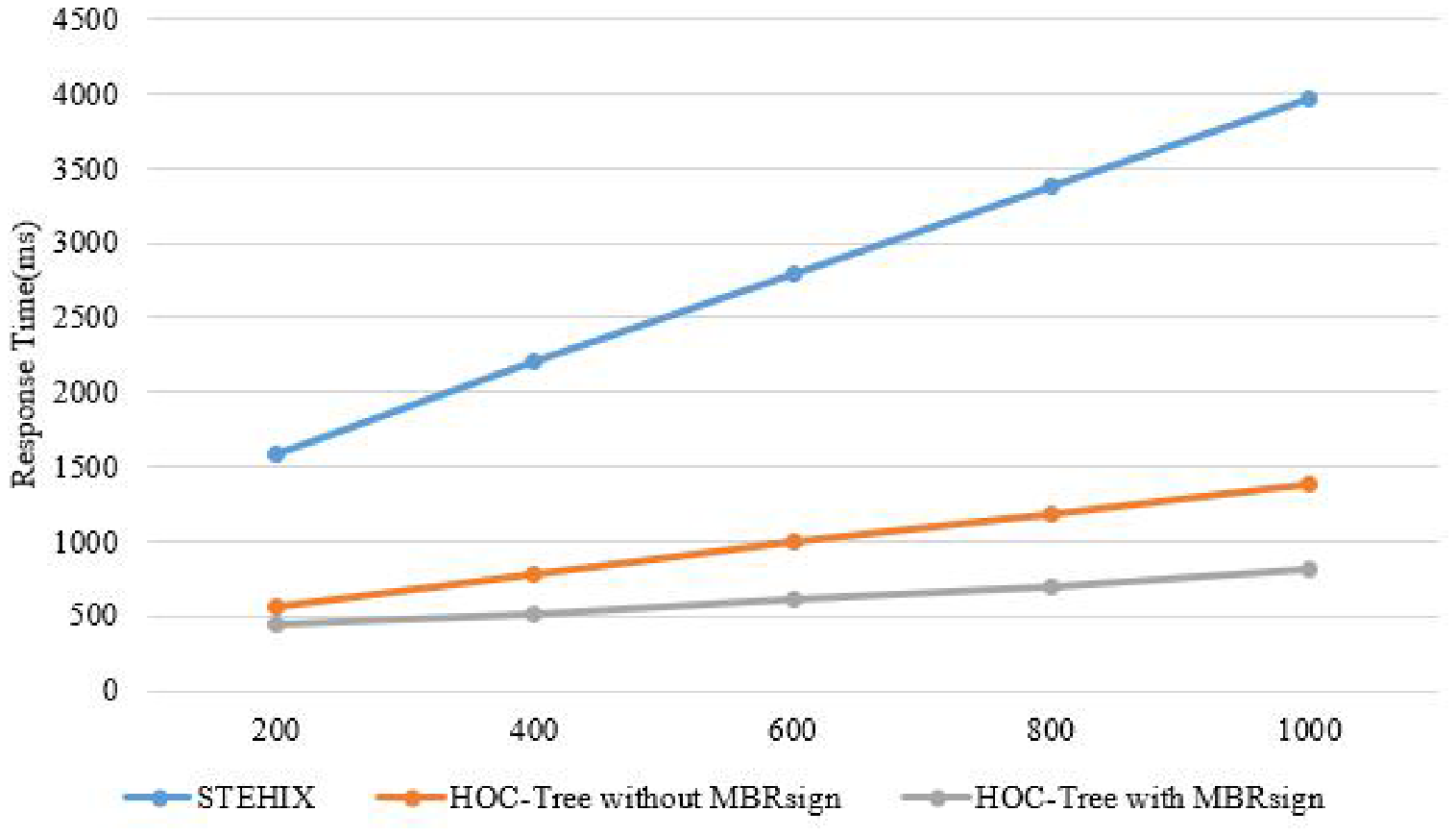}%
     }
   \captionsetup{justification=centering}
       \vspace{-0.2cm}
\caption{\small{Performance effected by temporal interval}}
\label{fig:Fig1516}
\end{center}
\end{minipage}
\label{fig:k}
\end{figure*}

To evaluate the effect of temporal interval on response time of HOC-Tree and STEHIX, experiments were conducted in the same manner with the previous one and spatial region was set as the default value 600. The experimental results are demonstrated in Fig.~\ref{fig:Fig1516}(a) and Fig.~\ref{fig:Fig1516}(b).

A large temporal query range would cover a lot of partial overlapping nodes, which fully satisfy with temporal restriction but non-fully satisfy with spatial restriction. As the temporal query range becomes larger in STEHIX, all these nodes have to be accessed by \emph{s-index} and \emph{t-index}, which increases the number of I/Os obviously and there are much more candidates to check in the refinement step while HOC-Tree has removed a lot earlier. Because, a \emph{MBRsign} tag data structure is designed in HOC-Tree reduce non-promising nodes access so that these nodes can be removed earlier to avoid unnecessary I/O costs. Fig.~\ref{fig:Fig1516}(a) and Fig.~\ref{fig:Fig1516}(b) demonstrate the running time of HOC-Tree with and without the tag, and our index performs better especially for skewed data. In such a scenario, the \emph{MBRsign} tag makes full use of non-fully-decoupled spatial and temporal information to confirm non-empty spatial covering nodes and thus many unnecessary I/O load can be avoided. For uniform dataset (UN), tag is still helpful when there are large number of empty partial overlapping nodes.

\section{Conclusions}
\label{concl}

The problem of spatio-temporal search is very significant due to the increasing amount of spatio-temporal data collected in widely applications. The proposed HOC-Tree is based on Hilbert curve and OC-Tree. Based on HOC-Tree, we design an efficient algorithm to solve the problem of spatio-temporal range search. The results of our experiments on real and synthetic data demonstrate that HOC-Tree is able to achieve a reduction of the processing time by 60-80\% compared with prior state-of-the-art methods.

\bibliographystyle{spmpsci}      

\bibliography{ref}

\end{document}